\documentclass[aps,prl,amsmath,amssymb,floatfix,twocolumn,superscriptaddress]{revtex4-1}
\usepackage{amssymb}
\usepackage{amsmath}
\usepackage[dvips]{graphicx}
\usepackage{mathrsfs}
\usepackage{ulem}
\usepackage[mathscr]{euscript}
\usepackage{gensymb}
\usepackage{color}
\usepackage{times}
\usepackage{dcolumn}
\usepackage{cases}
\usepackage{txfonts}
\usepackage{textcomp}
\usepackage[mathscr]{euscript}
\usepackage{indentfirst}
\usepackage{subfigure}
\usepackage{epstopdf}
\usepackage{graphicx}
\DeclareGraphicsExtensions{.png,.eps}

\begin{document}
\bibliographystyle{unsrt}	

\title{Thermodynamics of Spin-1/2 Kagom\'e Heisenberg Antiferromagnet: Algebraic Paramagnetic Liquid and Finite-Temperature Phase Diagram}

\author{Xi Chen}
\affiliation{School of Physical Sciences, University of Chinese Academy of Sciences, P. O. Box 4588, Beijing 100049, China}

\author{Shi-Ju Ran}
\affiliation{Department of Physics, Capital Normal University, Beijing 100048, China}
\affiliation{ICFO-Institut de Ciencies Fotoniques, The Barcelona Institute of Science and Technology, 08860 Castelldefels (Barcelona), Spain}
\author{Tao Liu}
\affiliation{School of Science, Hunan University of Technology, Zhuzhou 412007, China}

\author{Cheng Peng}
\affiliation{School of Physical Sciences, University of Chinese Academy of Sciences, P. O. Box 4588, Beijing 100049, China}

\author{Yi-Zhen Huang}
\affiliation{School of Physics and Astronomy, Shanghai Jiaotong University, Shanghai, 200240, China}

\author{Gang Su}
\email[Corresponding author. ] {Email: gsu@ucas.ac.cn}
\affiliation{School of Physical Sciences, University of Chinese Academy of Sciences, P. O. Box 4588, Beijing 100049, China}
\affiliation{Kavli Institute for Theoretical Sciences, and CAS Center for Excellence in Topological Quantum Computation, University of Chinese Academy of Sciences, Beijing 100190, China}

\begin{abstract}

Quantum fluctuations from frustration can trigger quantum spin liquids (QSLs) at zero temperature. However, it is unclear how thermal fluctuations affect a QSL. We employ state-of-the-art tensor network-based methods to explore the ground state and thermodynamic properties of the spin-$1/2$ kagom\'e Heisenberg antiferromagnet (KHA). Its ground state is shown to be consistent with a gapless QSL by observing the absence of zero-magnetization plateau as well as the algebraic behaviors of susceptibility and specific heat at low temperatures, respectively. We show that there exists an \textit{algebraic paramagnetic liquid} (APL) that possesses both the paramagnetic properties and the algebraic behaviors inherited from the QSL. The APL is induced under the interplay between quantum fluctuations from geometrical frustration and thermal fluctuations. By studying the temperature-dependent behaviors of specific heat and magnetic susceptibility, a finite-temperature phase diagram in a magnetic field is suggested, where various phases are identified.  This present study gains useful insight into the thermodynamic properties of the spin-1/2 KHA with or without a magnetic field and is helpful for relevant experimental studies.

\textbf{Keywords---} Kagom\'e antiferromagnet, Gapless quantum spin liquid, Algebraic paramagnetic liquid, Finite-temperature phase diagram.
\end{abstract}

\maketitle

\section{Introduction}

Quantum spin liquid (QSL) is an exotic state in which the interactions between spins fail to order the system at temperature even down to zero. Since it was suggested by Anderson \cite{QSL} forty years ago for a possible ground state of the triangular Heisenberg antiferromagnet, QSL has received much attention in condensed matter physics owing to its likely connection to high-temperature superconductor and topological phases in quantum magnetism. The search for a QSL has been attempted over past decades \cite{balents}. As a magnetic system with low spin, low dimensionality and strong frustration is usually thought to favor a QSL \cite{QSL},  the spin-$1/2$ Heisenberg antiferromagnet on kagom\'e lattice is widely viewed as the most promising candidate.
Nonetheless, after decades of intensive investigations, a mass of works reveal that, both theoretically and experimentally, the spin-$1/2$ kagom\'e Heisenberg antiferromagnet (KHA) should be a quantum spin liquid (QSL) \cite{KagomeQSL}, but the nature of its ground state, say, a gapless or gapped QSL, has still no consensus and is currently actively debated.

A number of theoretical studies on the spin-1/2 KHA by density matrix renormalization group (DMRG) \cite{DMRGgapped1, DMRGgapped2, DMRGgapped3, DMRGgapped4, DMRGgapped5}, symmetric tensor network state (TNS) \cite{WEN XG}, etc., tend to give a gapped $Z_2$ QSL. However, a few inconsistencies exist as well. For instance, the spinon excitations in a $Z_2$ QSL have not yet been observed. In addition, a chiral spin liquid (CSL) was also proposed to this model by adding second and third nearest-neighbor spin interactions \cite{CSL1, CSL2}. Since $Z_2$ QSL and CSL do not belong to the same universality class, possible transitions by turning on the second and third nearest-neighbor interactions need to be further investigated \cite{CT, Z2toCSL1, Z2toCSL2}. On the other hand, there are other works by e.g. variational Monte Carlo \cite{gapless1, gapless2, gapless3, gapless4, gapless5}, and various TNS based algorithms \cite{gapless8,gapless6}, favoring a gapless QSL. In particular, a recent large-scale DMRG simulation \cite{gapless7} finds indications for a gapless Dirac spin liquid. It is obvious that these studies make this issue quite subtle.

On the experimental aspect, the mineral Herbertsmithite $ZnCu_3(OH)_6Cl_2$ \cite{herbersmithite} is usually considered as an ideal model compound of the spin-$1/2$ KHA. Earlier inelastic neutron scattering and Raman spectroscopy \cite{Expgapless1, Expgapless2} as well as thermodynamic measurements \cite{herbersmithite, ExpThermo1,ExpMag} on Herbersmithite appear to support a gapless QSL, but a recent NMR study shows a finite spin gap around $0.03J \sim 0.07 J$ \cite{gappedNMR}, in favor of a gapped $Z_2$ QSL. The incompatibility in experiments also remains unsolved, albeit the Dzyaloshinskii-Moriya interaction is believed to play a role in physics of Herbertsmithite \cite{ExpMag,DM1,DM2,DM3}.

Besides the ground-state simulations, the finite-temperature simulations on KHA are equally important, which are, however, less implemented due to the lack of accurate and efficient methods. There are a few investigations by, e.g., the high-temperature series expansion \cite{HTSE1,HTSE2}, low-energy effective field theory \cite{LEF1,LEF2,M. Hermele}, projected-wave-function techniques \cite{M. Hermele,gapless1,gapless5}, and Hams–de Raedt method \cite{kagomeFT1,kagomeFT2}. These approaches are severely limited by finite sizes or the temperature/energy regime of interest. The finite-temperature simulations on KHA with tensor network (TN) methods \cite{ODTNS,NCD,FTPEPS,Variant_FTPEPS,FTTNR,FTTNR2}, which possess genuine advantages on capturing many-body features and have achieved great successes on the ground states, are still lack.

In this work, we utilize the TN-based numerical methods with two distinct optimization algorithms to study the spin-1/2 KHA with high accuracy. Firstly, we study the ground state and obtain the evidence favoring for a gapless QSL \cite{gapless1}. At low temperatures, the algebraic behaviors of magnetic susceptibility and specific heat are observed. At certain temperatures, the system is found to exhibit algebraic properties from the ground-state QSL but is still paramagnetic in the presence of a small magnetic field. We coin this phase as an algebraic paramagnetic liquid (APL). The APL can be understood as an intermediate thermal phase that connects the zero-temperature algebraic QSL to the high-temperature trivial paramagnetic phase. We speculate that the APL would generally emerge after introducing proper thermal fluctuations to a gapless QSL. 
By further studying the temperature-dependent behaviors of the specific heat and susceptibility in magnetic fields, a finite-temperature phase diagram is firstly suggested for this system. The phase diagram also reveals an interesting phenomenon that even when the QSL at zero temperature is frozen by the magnetic field to a solid state (e.g., a plateau phase), proper thermal fluctuations could ``melt'' the system into a liquid-like APL state.

\section{Model and Methods}

Let us consider the spin-1/2 Heisenberg antiferromagnet on kagom\'e lattice in a magnetic field. The Hamiltonian reads
\begin{equation}\label{eq1}
H = J \sum\limits_{<ij>}\textbf{S}_{i}\cdot\textbf{S}_{j} -h \sum\limits_{i}S^{z}_{i},
\end{equation}
where $J$ is the coupling constant, $<ij>$ stands for the summation over nearest neighbors, $\textbf{S}_i$ is the spin operator on the $i$th site, and $h$ is the magnetic field along the $z$ direction. We assume $g\mu_B=1$ ($g$ the Land\'e factor, $\mu_B$ the Bohr magneton) for simplicity.

We adopt two different tensor network \cite{Nishino01,Verstracte04,Verstracte04_2} methods such as iPEPS \cite{iPEPS} and ODTNS \cite{ODTNS} to study this system numerically. To contract the tensor networks of the ground state and the partition function, two update schemes, cluster update \cite{cluster1,cluster2,cluster3} and full update \cite{iPEPS,full-update2,full-update3}, are employed to optimize the truncation. In the cluster update algorithm, we choose the unit cell formed by 6 tensors that preserve the symmetry of the state and use the Bethe approximation to simulate the environment outside the cluster; in the full update algorithm, we contract the whole TN by iTEBD \cite{iTEBD}.

The errors of the TN algorithms come mainly from two aspects: the truncation error and the error of calculating the environment that determines the truncation. For example in the simple update, the environment is calculated by optimally ignoring all the loops in the original lattice. The cluster update considers all the loops inside the cluster. The full update provides a more accurate simulation of the environment by using iTEBD or CTMRG algorithms that approximately consider all the loops. It gives an accuracy higher than the simple/cluster updates when using the same $D_c$, the dimension cut-off of truncating tensors. But the full update is much more time-consuming, and thus, the bond dimension is usually limited to smaller values than in the cluster update. The simple/cluster update approximates the environment by the entanglement from a tree-like state, and can access much higher bond dimensions. More details can be found in a recent review about TN contractions \cite{TN-Review}.

At finite temperatures, we invoke the tensor product density operator (TPDO) \cite{ODTNS,NCD,FTPEPS,Variant_FTPEPS,FTTNR,FTTNR2} and calculate the relative quantities by means of the method proposed in Ref. \onlinecite{ODTNS}. Refer to Supplemental Material for the details of calculating algorithms. In ground state calculations, the Trotter step $\tau$ is taken from $10^{-1}$ to $10^{-5}$. We can reduce the Trotter error by reducing the Trotter slice, and here we only need the converged ground state. In simulations of thermodynamic properties, $\tau$ is fixed to be $10^{-2}$. In this case, we start with a high-temperature density operator at the inverse temperature $\tau$. Then in every step of the evolution, the TN represents the density operator at the corresponding temperature, and is used to calculate thermodynamic properties. Therefore, it is more convenient and efficient to keep $\tau$ unchanged during the simulation. In cluster and full update schemes, we choose a unit cell of nine spins that comprises of one hexagon and six triangles. The benchmark on the ground state energy is given in the Supplemental Material.

\begin{figure}[tbp]
	\includegraphics[width=1\linewidth]{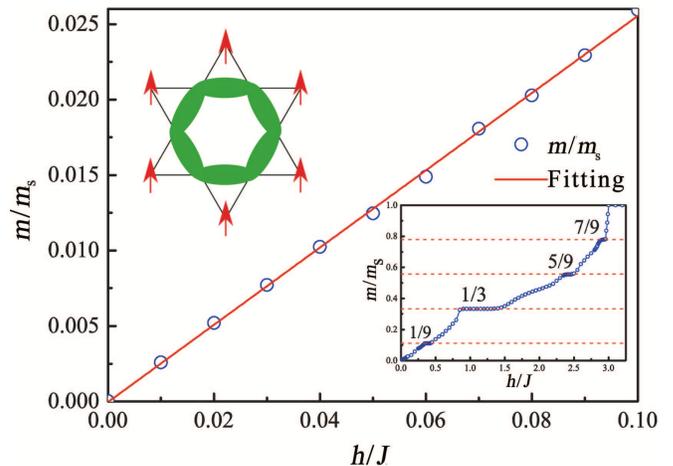}
	\caption{(Color online) The magnetization per site $m$ versus magnetic field $h/J$ at low fields. The magnetization vanishes linearly when $h/J$ tends to zero.  Lower inset presents the overall magnetic curve, where the magnetization plateaus at $m/m_s=1/9, 1/3, 5/9, 7/9$ (with $m_s$ the saturated magnetization per site) are clearly observed, but zero magnetization plateau is absent. Upper inset gives the spin configuration of the 1/3-magnetization plateau state, in which a singlet hexagon valence bond state (bold green hexagon) forms with the three spins at the corner (In our numerical method, we use the translational invariance. Each spin at the outer corner is shared by two clusters, so for each cluster, it contains 9 spins) almost fully polarized. These results are calculated by the cluster update algorithm with $D_c = 10$.}
	\label{fig1}
\end{figure}

\section{Gapless quantum spin liquid}
The gapless QSL in the ground state is supported by the behaviors of ground-state magnetization, the low-temperature specific heat and susceptibility, respectively. Fig. \ref{fig1} presents the ground-state magnetization per site $m$, defined by $m=(1/N)\sum_{i}\langle S_{i}^{z} \rangle$ (with $N$ the total number of lattice sites) as a function of the magnetic field $h$. It can be seen that $m$ depends linearly on $h$ at small magnetic fields, i.e., $m \sim h^{\alpha}$ with $\alpha=1$, and vanishes as $h \rightarrow 0$. These results are in accordance with a U(1) instead of Z$_2$ QSL, consistent with the projected wave function study \cite{gapless1} and a recent simulation by DMRG on infinite long cylinders \cite{gapless7}. Note that previous DMRG calculations on cylinders \cite{DMRGgapped1,DMRGgapped2,DMRGgapped3,DMRGgapped4,DMRGgapped5,plateau1} claimed a gapped $Z_2$ spin liquid with a small gap on this model. There was also a recent work based on the tensor entanglement renormalization group with symmetry, giving a gapped ground state \cite{WEN XG}. The controversies might be from different restrictions and biases of these algorithms. For instance, DMRG is only applied to cylinders and finite-size clusters; and the accuracy of tensor network methods relies on various update schemes. Hence, it is strongly demanded to make more investigations other than simulating the ground state.


In addition, we calculate the overall magnetic curve shown in the lower inset of Fig. \ref{fig1}. One may see that there are four plateau states occurring at $m/m_s = 1/9, 1/3, 5/9, 7/9$, with $m_s$ the saturated magnetization per site. The non-zero plateaus are consistent with the DMRG calculations in Ref. \onlinecite{plateau1} and the ED calculations in Ref. \onlinecite{plateau2}. Note that an important difference is that a narrow zero plateau was observed in these works, which might be due to the finite-size effects. Surely, further DMRG or ED simulations are still needed. The non-zero plateaus can be written in a compact form of $m/m_s=(1-2\ell/9)$, $\ell=1,2,3,4$. For the $m/m_s=1/3$ plateau state, we observe a ferrimagnetic order as shown in the upper inset of Fig. \ref{fig1}. It is interesting to note that the $m/m_s=1/9$ plateau was observed in a previous DMRG calculation on cylinders, where this 1/9-plateau was claimed to be a $Z_3$ spin liquid \cite{plateau1}. It is also worth mentioning that for the $m/m_s=1/3$ plateau state, the six spins on the hexagon of the unit cell form a singlet hexagon valence bond state with vanishing magnetic moment on each site, while the outer three spins are almost fully polarized \cite{plateau1,plateau2} [the upper inset of Fig. \ref{fig1}]. This is in contrast to an up-up-down spin configuration observed on an infinite Husimi lattice \cite{gapless8}.

\begin{figure}
    \begin{minipage}{\linewidth}
        \includegraphics[width=1\linewidth]{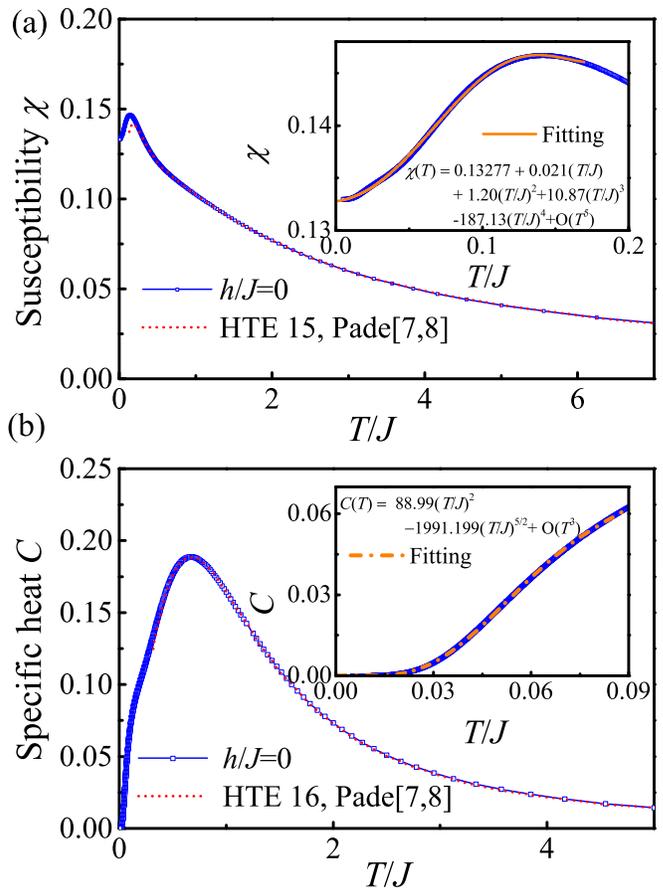}
    \end{minipage}
\caption{\label{fig2}
(Color online) The susceptibility $\chi$ (a) and specific heat $C$ (b) as a function of temperature $T$ for the spin-$1/2$ kagom\'e Heisenberg antiferromagnet in the absence of magnetic field. Here the cluster update algorithm with bond dimension $D_c = 17$ is adopted. $\chi$ (blue open circle in (a)) and $C(T)$ (blue open square in (b)) are compared with those of the fifteenth-order high temperature series expansion (HTE) (red dash line in (a)) and the sixteenth-order HTE (red dash line in (b)) under parameters Pade[7,8], respectively. As shown in the insets, both $\chi$ and $C$ exhibit algebraic behaviors at low temperatures. These results are calculated by the cluster update algorithm with $D_c = 17$.
}
\end{figure}

Besides the absence of zero magnetization plateau, more solid evidences of the gapless QSL are obtained from the algebraic behaviors of the temperature-dependent susceptibility $\chi =\partial m/ \partial h$ and specific heat $C = \partial U / \partial T$ (with $U$ the internal energy) at low temperatures. Fig. \ref{fig2} (a) shows the temperature-dependence of $\chi(T)$ in zero magnetic field ($h=0$). At high temperatures, our results coincide with the fifteenth-order HTE calculation \cite{HTSE1,HTSE2} under the parameter Pade [7,8]. For $h=0$, $\chi(T)$ can be nicely fitted by a polynomial $\chi (T)= 0.13277 + 0.02096(T/J) + 1.20337(T/J)^2 + O(T^3)$ at low temperatures [the inset of Fig.~\ref{fig2} (a)]. When $T \rightarrow 0$, $\chi(T)$ approaches {\it linearly} to a constant $0.13277$ which is consistent with the ground state susceptibility $\chi(0) = 0.1298$ given by the slope of the magnetic curve at small $h$ (Fig. \ref{fig1}). This temperature dependent behavior of the susceptibility resembles the result from the modified spin-wave theory \cite{takahashi}, where the linear $T$-dependence of $\chi$ at $T \rightarrow 0$ is obtained for the spin-1/2 Heisenberg antiferromagnet on square lattice, which has gapless excitations (Goldstone modes) because it has an antiferromagnetic long-range ordered ground state. However, we do not find any long-range order in the spin-1/2 KHA, suggesting that the spin excitations in this system could be essentially different from the square Heisenberg antiferromagnet.

For the specific heat $C$, a peak appears in the absence of magnetic field as demonstrated in Fig.~\ref{fig2} (b). The specific heat $C$ at high temperatures obtained by TN scheme also coincides with the results from the sixteenth-order HTE with the parameter Pade [7,8] \cite{HTSE1,HTSE2}. At low temperatures, $C(T)$ can be well fitted by a polynomial function $C(T) = 88.991(T/J)^2-1991.199(T/J)^{5/2} + O(T^3)$  [inset of Fig. \ref{fig2} (b)]. It suggests that, when $T\rightarrow 0$, we have $C(T) \sim T^2$, which is another evidence for the gapless QSL \cite{Ran}.


\begin{figure}
	\includegraphics[angle=0,width=1\linewidth]{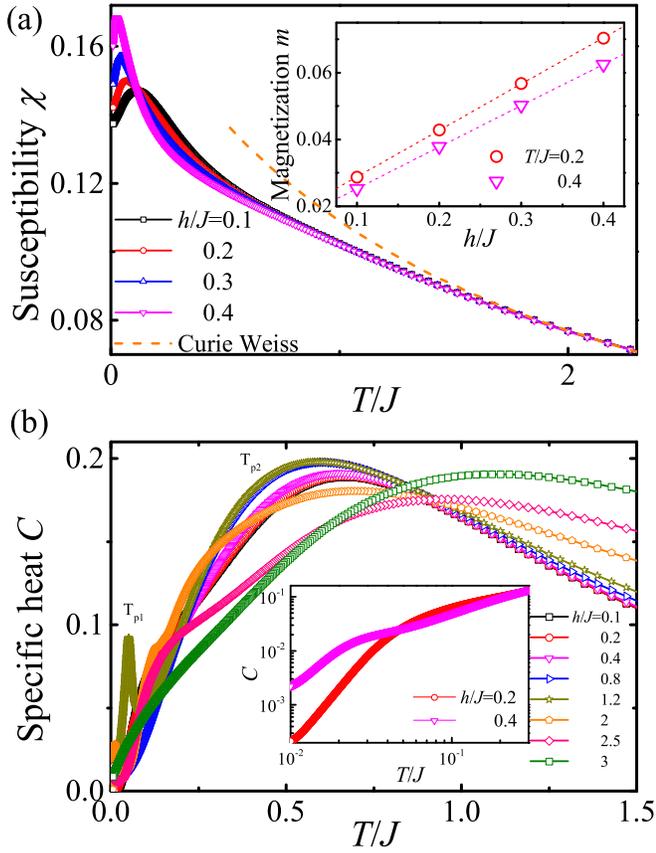}
	\caption{(Color online) Temperature dependence of (a) susceptibility $\chi (T)$ and (b) specific heat $C(T)$ of the spin-$1/2$ kagom\'e Heisenberg antiferromagnet for various magnetic fields $h$. At high temperatures, the susceptibility obeys the Curie-Weiss paramagnetic law. When temperature goes down to the APL region, it shows a paramagnetic behavior of the linear dependency of magnetizaion on $h/J$ [inset of (a)]. The positions of round peaks of $C(T)$ almost appear at the same temperature for small $h/J$, but grow linearly with $h$ when $h/J > 1.5$. In the APL phase, $C(T)$ shows an algebraic behavior [a logarithmic plot in the inset of (b)].  These results are calculated by the cluster update algorithm with $D_c = 17$.}
	\label{fig3}
\end{figure}

\section{Algebraic paramagnetic liquid and finite-temperature phase diagram}

Under small $h$, we find that the susceptibility $\chi$ and specific heat $C$ preserve the algebraic properties at low temperatures (Fig. \ref{fig3}), similar to the ground-state QSL found at $h=0$. It is clear that at high temperatures, it obeys the Curie-Weiss law [$\chi^{-1} \propto (T/J+1.4)$]. The paramagnetism at low temperatures is unveiled by the linearly field-dependent magnetization, as shown in the inset of Fig. \ref{fig3} (a). Hence, we dub this finite-temperature phase that possesses both paramagnetic and the liquid-like algebraic properties as APL. These algebraic properties indicate that the APL is a liquid-like state at finite temperatures that essentially differs from the conventional paramagnetic phase. APL can be understood as an intermediate thermal phase that connects the zero-temperature algebraic QSL to the high-temperature trivial paramagnetic phase. The algebraic behavior of APL is inherited from the algebraic QSL ground state.

\begin{figure}[tbp]
	\includegraphics[angle=0,width=1\linewidth]{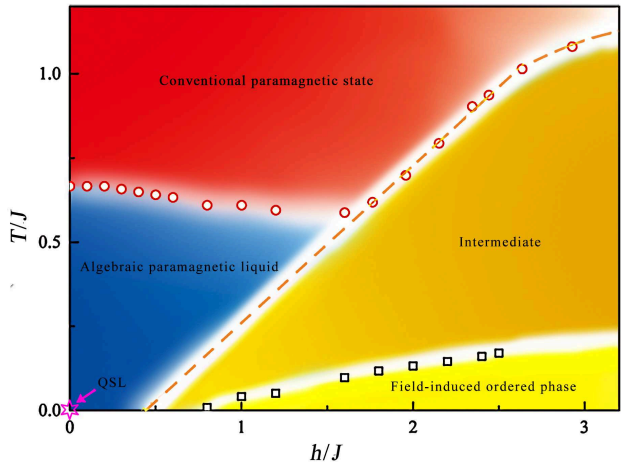}
	\caption{(Color online) The phase diagram of the spin-1/2 KHA in the plane of temperature and magnetic field, which consists of the phases including gapless quantum spin liquid (QSL), algebraic paramagnetic liquid, field-induced ordered phase, intermediate (spin canted) phase, and conventional paramagnetic state. The upper (red circles) and lower (black squares) phase boundaries are determined from the high- and low-temperature peaks of the specific heat in different magnetic fields, respectively. The dash line is obtained by fitting the upper boundary for large fields.}
	\label{fig-Phase}
\end{figure}

We give the phase diagram in the plane of temperature and magnetic field as shown in Fig. \ref{fig-Phase}. The phase boundary at relatively higher temperatures (denoted by $T_{p2}$, red circles) corresponds to the high-temperature peak of $C(T)$. One can see two different behaviors of $T_{p2}$ when $h$ changes. For $h/J < 1.5$, $T_{p2}$ changes very little with $h$. The peak of $C(T)$ in this region is dominated by the low-energy physics inherited from the QSL, and the insensitivity of its position $T_{p2}$ to $h$ is consistent with the fact that the ground-state QSL has no magnetic momentum to gain or loss energy when $h/J$ changes. Thus, this line of $T_{p2}$ could be regarded as the signature that separates the APL phase from the trivial paramagnetic phase at high temperatures. Since both the APL and paramagnetic phases have the same symmetry, there is no conventional phase transition but a crossover between this two phases.

Moreover, a universal scaling for the specific heat appears for $h/J < 1.5$. The scaling equation, $C(1+h/J)^\alpha = q_0 + b_1(1+h/J)^\beta (1-\frac{T}{T^*})$, is found in the vicinity of an isosbestic point $T^*/J=0.25126$. Refer to Supplemental Material for details. This implies that the phase below this upper boundary should be dominated by the low-energy physics of the QSL, which gives rise to a new phase dubbed as the APL. Above this boundary, we have the high-temperature trivial paramagnetic phase.

For about $1.5<h/J<2.5$, $T_{p2}$ increases linearly with increasing $h$, i.e. $T_{p2} = \alpha h + const$ with $\alpha \simeq 0.45$. Note for the noninteracting spin model $\hat{H} = h\sum_i \hat{S}^z_i$, the cross-over temperature also satisfies a linear relation with $h/J$ with a similar slope as $\alpha^\prime \simeq 0.43$. The difference on the constant interceptions might be caused by the occurrence of independent magnons. A direct comparison is given in the Supplementary Material. For about $h/J>2.5$, a change of the interception occurs and then $T_{p2}$ keeps to grow linearly with the same slope. These suggest that the crossover occurring in this region is dominated by the high-temperature trivial paramagnetic states.

We shall stress that for $h/J<1.5$, the contribution with the noninteracting nature should still exist. Since the QSL dominates here, we cannot directly distinguish in this region the peak of $C$ for $h/J>1.5$. For this reason, we extrapolate the boundary according to the linear relation to $T=0$ (dash line). This should separate the APL and intermediate phases. Note that at $T=0$, the intercept of the extrapolation is close to the left boundary of $1/9$ plateau.

For the intermediate (spin-canted) phase, it normally appears between the high-temperature paramagnetic phase and low-temperature ordered phase of 2D models. The boundary between the intermediate and the field-induced ordered phases is given by the positions of the low-temperature peak of $C(T)$ (denoted by $T_{p1}$, see black squares in Fig. \ref{fig-Phase}). This boundary ends at $h/J \simeq 0.8$ in the $T \rightarrow 0$ limit, which coincides with the left boundary of the $1/3$ plateau phase. In the field-induced ordered phase ($T<T_{p1}$), the magnetization per site remains nonzero. When $T>T_{p1}$, the thermal fluctuations tend to destroy the ordering and restore the broken symmetries, driving the system into a spin-canted (intermediate) phase. As $T$ continues to rise, the canted spins will fully restore the symmetries and smoothly cross over to the trivial paramagnetic phase. Meanwhile, the magnetization eventually vanishes for $T>T_{p2}$.

Since the system does not spontaneously break any local symmetries at $T = 0$, we observe no conventional phase transitions at finite temperatures. Note that the finite-temperature transitions/crossovers of QSL models are still unclear currently. The phases are distinguished by physical properties (e.g., peaks of specific heat, susceptibility, etc.), and are connected by cross-overs according to our results. It is also possible that the transitions might be described by nonlocal order parameters (e.g., topological orders). Current techniques are not able to identify this particularly at finite temperatures.

Our work poses an interesting phenomenon. Starting from $h=0$ and $T=0$, where the strong quantum fluctuations permit a disordered QSL, the system can be frozen into a solid-like state by increasing the magnetic field to, e.g., $h/J=1$. The reason is that a larger field suppresses the frustration, reducing the quantum fluctuations. Then by properly increasing temperature to, e.g., $T/J = 0.5$, the system is melted by thermal fluctuations, and enters into the APL state that possesses nontrivial QSL-like properties. Thus, the APL is the consequence of interplay between quantum and thermal fluctuations. We speculate that the APL phase could generally emerge after introducing proper thermal fluctuations to a gapless QSL.



The finite-temperature data given in our work can be easily compared with experiments. It is interesting to remark that our phase diagram is similar to the one suggested by the $^{17}O$ NMR measurements in Herbertsmithite in a magnetic field \cite{ExpMag}. Both show an algebraic (QSL and APL) phase and a solid-like (spin solid or intermediate) phase with similar phase boundaries. More efforts should be done in future to push numerical calculations to the whole temperature region in experiments. 

\section{Summary}

By utilizing state-of-the-art tenser network algorithms, we show that the ground state of the spin-$1/2$ KHA is a gapless QSL, which is evidenced by the absence of zero-magnetization plateau, as well as the algebraic behaviors of susceptibility and specific heat at low temperatures. By studying the effects of thermal fluctuations on the QSL, we obtain the phase diagram in the $h$-$T$ plane, where the five phases are identified, which are QSL, APL, field-induced ordered state, canted (intermediate) phase, and high-temperature trivial paramagnetic phase. The APL phase is unveiled to emerge from both quantum and thermal fluctuations, which possesses paramagnetic behaviors (the linear field-dependence of magnetization) and the QSL-like algebraic properties observed from the magnetic susceptibility and specific heat. This present work indicates that even when the QSL at zero temperature can be frozen into a solid state by adding a magnetic field, proper thermal fluctuations could ``melt'' the system again into a liquid-like APL state. In addition, the relevance of our phase diagram for the spin-$1/2$ KHA to the one suggested experimentally for Herbertsmithite compound in a magnetic field is also addressed. As the thermodynamic calculations on the spin-1/2 KHA in a magnetic field are still sparse, our findings would spur more experimental explorations on this fascinating system, and shed deeper insight on the novel states of matter.

\acknowledgments
The authors acknowledge Wei Li and Xin Yan for useful discussions. This work was supported in part by the MOST of China (Grant No. 2018YFA0305800), the NSFC (Grant No. 14474279, 11834014), and the Strategic Priority Research Program of the Chinese Academy of Sciences (Grant No. XDB07010100, XDPB08). SJR was supported by ERC AdG OSYRIS (ERC-2013-AdG Grant No. 339106), Spanish Ministry MINECO (National Plan 15 Grant: FISICATEAMO No. FIS2016-79508-P, SEVERO OCHOA No. SEV-2015-0522), Generalitat de Catalunya (AGAUR Grant No. 2017 SGR 1341 and CERCA/Program), Fundaci\'o Privada Cellex, EU FETPRO QUIC (H2020-FETPROACT-2014 No. 641122), the National Science Centre, and Poland-Symfonia Grant No. 2016/20/W/ST4/00314, and Fundaci\'o Catalunya - La Pedrera $\cdot$ Ignacio Cirac Program Chair.

\end{document}